\documentclass[12pt]{article}
\usepackage[a4paper,margin=1in]{geometry}
\usepackage[square,numbers,sort&compress]{natbib} 
\usepackage{indentfirst}
\usepackage{amsmath} 
\usepackage{graphicx}
\usepackage{amssymb}
\usepackage{booktabs}
\usepackage{subcaption}
\usepackage{tikz}
\usetikzlibrary{arrows.meta, positioning, shapes.geometric}
\usepackage{multirow}
\usepackage{tikz}
\usetikzlibrary{calc}
\usetikzlibrary{decorations.pathreplacing}  
\usetikzlibrary{arrows.meta}                
\usepackage{tikz}
\usetikzlibrary{calc,decorations.pathreplacing,arrows.meta}
\usepackage{tikz}
\usetikzlibrary{positioning}
\usepackage{float}
\usepackage{indentfirst}
\title{Random Forest-Informed Cellular Automaton for Large-Scale Wildfire Spread Modelling}
\author{
Siyu Chen$^{1}$,
Esha Saha$^{2}$,
Hao Wang$^{3}$
}

\date{
$^{1}$First author\\
$^{2}$Second author\\
$^{3}$Corresponding author
}

\date{}

\begin{document}

\maketitle

\begin{abstract}
Accurate large-scale wildfire spread modelling requires models that capture both the environmental conditions associated with fire occurrence and the local dynamics of fire propagation. We propose a three-stage framework that combines a Random Forest (RF) model with a cellular automaton (CA). First, an RF model trained on the 2021 Canadian fire season estimates daily pixel-level fire-occurrence probabilities. Second, quantile gradient boosting models provide optional spread-rate priors for sensitivity analysis. Third, an RF-informed CA combines the RF probability layer with neighbourhood-driven spread on a 5 km grid. The RF model achieved AUC values of 0.725--0.795 on the 2022--2024 datasets, while the RF-informed CA achieved substantially higher spatial overlap than the evaluated CA-only baselines in the 2023 simulation. A higher-resolution simulation provides an additional qualitative assessment of local spatial errors. These results suggest that combining RF-derived probabilities with local CA spread can improve large-scale wildfire simulations under the tested conditions.
\end{abstract}

\section{Introduction}

Wildfires have intensified across Canada in recent decades, driven by
accelerating climate change, extreme weather events, and expanding
wildland-urban interfaces. The 2023 fire season, which produced record burned area and severe smoke impacts across North America \cite{NRCan2024RecordWildfires,Jain2024Drivers2023Canada}, highlighted the urgent need for reliable tools that can anticipate fire ignition and large-scale spread. As national fire management increasingly depends on data-driven decision systems, developing models that remain stable under heterogeneous, noisy, and operationally imperfect datasets has become a central scientific and practical challenge.

Wildfire research encompasses various approaches such as fire danger assessment, fire
occurrence prediction, fire spread simulation, and fire effects analysis, to name a few. This study focuses on two modeling tasks that are often studied separately in literature but are closely connected in practical applications, namely fire occurrence prediction and fire spread simulation. 
Fire occurrence prediction for example, is
mainly used for fire danger assessment and early resource planning, while fire spread simulation is used to evaluate how fires may develop over a spatio-temporal domain.
Fire occurrence prediction mainly uses statistical and machine learning models to estimate the risk of fire ignition based on meteorological, topographic, vegetation, and human related variables.
Methods for fire spread modelling vary from solving mathematical models to using data-driven methods such as random forests and neural networks \cite{Jain2020Review, Naderpour2021Forest, Kalantar2020Forest,srivastava2026spatiotemporalwildfirespreadprediction, saha2025learningcoupleddynamicsincomplete}.
Cellular automata and related rule based models \cite{Karafyllidis1997, Encinas2007Hexagonal, Encinas2007Fronts} are widely used to simulate fire spread because of their clear structure and explicit spatial form. However, these models often rely on
empirically defined spread parameters, which can reduce their stability when
applied to noisy data or across different years \cite{Karafyllidis1997,Alexandridis2008,Zheng2017,Xu2022}.

In existing studies that combine machine learning with cellular automata for wildfire spread modelling, a common approach is to embed machine learning components directly within the spread mechanism itself \cite{Zheng2017, Xu2022, Li2022LSTMCA}. In such frameworks, machine learning models are used either to approximate the fire spread dynamics\cite{Li2022LSTMCA,Wu2022,Cheng2022,Zohdi2020}    or to determine the transition rules governing cellular automata behaviour\cite{Zheng2017, Xu2022}.
For example,\cite{Kozik2013} proposed a fire spread model based on an artificial neural network combined with Kalman filtering for data assimilation, where the resulting structure functionally resembled a complex cellular automaton rather than a traditional neural network. 
Similarly, \cite{Zheng2017}  integrated a cellular automata model with an extreme learning machine (ELM), in which the ELM was trained on historical fire, vegetation, topographic, and meteorological data to directly determine the transition rules of the cellular automaton.
These approaches reduce reliance on manually specified rules and demonstrate the potential of data-driven methods for modelling fire spread processes. 
However, such methods depend heavily on the specific design of the cellular automaton, including the neighborhood structure, transition rules, and spatial resolution.
As a result, changes in spread rules, spatial resolution, or input variables often require retraining or redesign of the machine learning component, increasing model complexity and limiting portability across regions and data conditions \cite{Jain2020Review}.
\textcolor{black}{In contrast, this study uses an RF-informed cellular automaton framework. The Random Forest works as an independent daily fire-occurrence model and produces daily probabilities for each location. The cellular automaton combines these probabilities with local neighbour-driven spread.}
The two components are connected through clearly defined inputs and outputs, without directly modifying the spread rules. The RF and neighbour-spread probabilities are combined using a noisy-OR operation, so the RF layer informs the simulation rather than acting as a hard constraint. \textcolor{black}{In this design, the Random Forest identifies areas where burning is more likely, while the cellular automaton adds spatial spread through local neighbour rules. Within this framework, the spread parameter $\alpha$ is not the main modelling target. It is used only to adjust the strength of neighbour-driven spread in the CA. Therefore, testing different $\alpha$ settings helps us understand how local spread changes the RF-informed simulation. The goal is not to find a universally optimal $\alpha$, but to evaluate the behaviour of the RF-informed CA framework.} This modular design keeps the roles of the RF and CA components interpretable.

\section{Related Work}

Wildfire modeling across Canada faces fundamental challenges arising from multi-source heterogeneity, incomplete observations, and strong spatio-temporal non-stationarity. Prior works consistently highlight the limitations that undermine the robustness and generalization ability
of both data-driven and mechanistic modeling frameworks
\cite{Jain2020Review, Ghali2023DLFire}. In this section, we discuss some challenges in data-driven wildfire modeling and relevant literature that has contributed to ignition modeling and wildfire spread.


\subsection{Ignition Modeling}
Early wildfire ignition modeling primarily relied on statistical approaches such as logistic regression and generalized linear models (GLMs), which estimate ignition probability as a function of weather, fuel conditions, topography, and human activity \cite{Chang2013,Vilar2011}. 
These models have been widely used because of their interpretability and clear statistical structure. However, their ability to capture complex nonlinear interactions among predictors is limited, which has motivated the increasing use of machine learning approaches in recent studies.
Machine learning (ML) methods, particularly Random Forests (RF), Gradient
Boosting models, and deep neural networks have achieved strong performance
in ignition prediction across diverse environments
\cite{Ghorbanzadeh2019, Pang2022}. RF has been
particularly influential due to its robustness against noisy inputs and
ability to model nonlinear interactions among weather, topography, and
fuel characteristics. 
Reviews consistently show that ML approaches outperform traditional statistical models in heterogeneous landscapes such as Canada \cite{Jain2020Review,Ghorbanzadeh2019,Pang2022,Naderpour2021Forest}. 

However, existing work also identifies three persistent limitations.
\textcolor{black}{First, most pixel-based machine-learning models for fire occurrence or ignition probability estimate the likelihood of ignition at each cell, but they do not directly simulate spatial continuity or neighbour-driven fire growth. 
This creates a gap between ignition prediction and spread modelling \cite{Jain2020Review,Shmuel2022GlobalWildfireSusceptibility}. 
Models that explicitly represent fire spread usually need to be coupled with cellular automata or process-based spread simulators \cite{Wu2022,Zohdi2020}.}
\textcolor{black}{Second, regression-based ML approaches can struggle with extreme spread behaviour and may underestimate large fire events. 
Third, ML models are sensitive to class imbalance and noisy or uneven spatio-temporal sampling, which are common challenges in wildfire occurrence prediction \cite{Jain2020Review,Illarionova2025GeoSpatialMLWildfire}.}
\textcolor{black}{For these reasons, RF is useful as an ignition-risk mapping tool, but it cannot replace a spatially explicit spread model such as a Cellular Automaton (CA).}

\subsection{Cellular Automata for Wildfire Spread}
\textcolor{black}{Existing wildfire spread models are traditional
physics-based and operational fire growth models, based on reaction diffusion equations that describe fire propagation using partial
differential equations (PDEs) representing heat transfer and fuel consumption processes. 
In operational settings, the Canadian PROMETHEUS model is widely used to simulate fire growth \cite{Tymstra2010Prometheus}. 
The model is based on the Canadian Fire Behaviour Prediction (FBP) System \cite{ForestryCanada1992FBP} and links spread rates to weather, fuel, and topographic conditions. Although these models provide interpretable predictions, they often require detailed fuel information and calibration, which limits their application to large scaled datasets.
}

A class of well-known methods in literature for simulating fire spreads are Cellular Automata (CA) models. 
They have been widely used for wildfire spread simulations due to their interpretability, computational efficiency, and explicit spatial formulation \cite{Karafyllidis1997, Alexandridis2008}.
They encode fire propagation through local transition rules within a
neighbourhood structure, making them attractive for operational use and
scenario analysis.
However, traditional CA models have several limitations.
First, key parameters such as the spread coefficient $\alpha$ and transition probability in CA models lack statistical grounding and require manual tuning, resulting in extreme sensitivity to initial conditions and environmental noise \cite{Karafyllidis1997,Alexandridis2008}.
These
issues are amplified when applied to noisy or interpolated datasets such as CFSDS.
Second, neighbourhood structures in classical CA models (Moore, von Neumann) \cite{Karafyllidis1997,Alexandridis2008,Encinas2007Hexagonal} are local by design and may simplify nonlocal processes such as ember transport, spotting, or wind-driven jumps across natural barriers \cite{Wadhwani2022FirebrandReview}.
Despite some extensions of CA model\cite{Karafyllidis1997,Alexandridis2008,Encinas2007Hexagonal,Wadhwani2022FirebrandReview}
incorporating wind and fuel moisture, structural limitations, especially for
large or fast-moving fires remains.
Third, CA models often generalize poorly across years, since parameters calibrated for one season typically fail in another due to nonstationary weather-fuel interactions. 
Absence of a principled calibration framework restricts CA applicability for
continental-scale modeling.
While hybrid models combining ML with CA, (such as CA coupled with Extreme
Learning Machines or latent data assimilation) provide partial improvements
\cite{Zheng2017, Cheng2022}, they do not fully resolve the mismatch between
CA assumptions and the complex realities of Canadian wildfire behaviour.






These insights suggest a need for stabilizing hybrid models by leveraging
the predictive strengths of ML with CA models to explicitly account for data
imperfections.
\textcolor{black}{Accordingly, the present study does not treat $\alpha$ as a stand-alone fitted objective. Instead, the RF provides a daily spatial probability field, while the CA models local neighbour-driven propagation. The two components are combined probabilistically in the CA update.}

\subsubsection*{Contributions of This Study}
\textcolor{black}{This study proposes an RF-informed CA framework for large-scale wildfire spread modelling. 
The Random Forest provides a daily fire-occurrence probability layer, while the CA models local neighbour-driven propagation. 
This design separates the RF probability layer from the local spread rules, making the model easier to interpret and adjust. Stage~2 provides optional quantile-based spread-rate scenarios for sensitivity analysis.} 


\section{Methodology}

In this section, we provide a detailed description of the proposed methodology, a three-stage framework for estimating daily fire occurrence and simulating wildfire burn progression.
We first outline the study area and data processing steps, followed by an explanation of all the stages in the framework.

\subsection{Data Collection and Processing}
We focus on wildfire activity across Canada, where different types of data were collected from sources such as satellite imagery and weather stations.
\textcolor{black}{For computational efficiency, the national dataset was partitioned into ten longitudinal slices, and slices 4–6 were selected as the focus region for model development and evaluation.}
The main sources of data were the Canadian Fire Spread Dataset (CFSDS) \cite{Barber2024CFSDS} and the National Burned Area Composite (NBAC) dataset \cite{Hall2020NBACCanada,Skakun2022ExtendingNBAC}.
The CFSDS consists of daily fire progression data, and the corresponding environmental covariates, describing the conditions on the day it burned. 
These covariates cover a range of factors: fire weather and drought indices (the six standard components of the Canadian FWI System derived from ERA5-Land meteorological reanalysis), local weather variables (temperature, humidity, wind, and precipitation on that day), topography (elevation, slope, aspect, and related indices derived from ASTER GDEM), forest fuel and vegetation attributes (from the national SCANFI data), and human influence indicators (e.g., road density and distance to roads, calculated from national transportation datasets at multiple scales).
\textcolor{black}{A summary of the predictor variables used in our models, as well as the excluded covariates and their rationale, is provided in the Supplementary Material.}
Fire event perimeters and ecozone information were obtained from the NBAC dataset. It also provides annual wildfire polygons with attributes such as start date, end date, and the affected ecozone, which we used for the framework analysis.
\textcolor{black}{The CFSDS pixel-level daily fire progression data is derived from satellite observations. 
Further details on data construction are provided in the Supplementary Material.}

\subsubsection{Data Preprocessing}

The CFSDS data have a spatial resolution of $180~\mathrm{m}$, with each pixel representing a $180~\mathrm{m} \times 180~\mathrm{m}$ area. the data was resampled to match the 180 m spatial resolution (using mean aggregation for the continuous variables and mode for the categorical variables). 
We filtered fire detections with spatiotemporal constraints based on NBAC fire perimeters. 
A hotspot detected by satellite was kept only if it occurred within 1 km of a known fire’s boundary and within a 30-day window before the fire’s reported start date or end date. We removed any low-confidence detections by \textcolor{black}{following the filtering criteria used in the CFSDS dataset, where MODIS hotspots with confidence $\leq 10$ (range 0–100) and VIIRS hotspots classified as \textit{LOW} are excluded, as these detections are commonly associated with sun glint or other noise sources \cite{Barber2024CFSDS}.} 
Further, all detection times were converted to local standard time to account for time zone differences.
In order to focus on the most active period, we restricted our dataset to the peak fire seasons corresponding to calendar days 150 to 250. This filtering was applied consistently to years 2021 through 2024, enabling strict out-of-year evaluations.
We treat data for the year 2021 for training and years 2022-2024 for validation/testing.
The final dataset is organized as a table of pixel-day records. 
Each record corresponds to one pixel on one day, with columns representing the predictors and response variable.
Missing data was handled using median imputation where for each variable, we calculated the median (for the training and validation sets separately) and use it to fill missing entries.

\subsection{The Three-Stage Model}

\begin{figure}[H]
\centering
\hspace*{-0.9cm} 
\begin{tikzpicture}[
    node distance=1.2cm and 1.0cm, 
    every node/.style={
        draw,
        rectangle,
        rounded corners,
        align=center,
        font=\small,
        minimum width=4.0cm,  
        minimum height=1.05cm
    },
    arrow/.style={->, thick}
]

\node (data) {Wildfire Data\\
CFSDS, NBAC, Meteorology, FWI, Terrain};

\node[below left=of data, xshift=0cm] (stage1) {Stage 1: Daily\\Fire-Occurrence Model\\
Random Forest\\
$p_i^{\mathrm{RF}}(t)$};

\node[below right=of data, xshift=-0cm] (stage2) {Stage 2: Spread-Rate Model\\
Quantile GBM\\
$\alpha(t;q)$};

\node[below=4.2cm of data] (stage3) {Stage 3: Cellular Automaton\\
Spatial Propagation (5 km grid)};

\node[below=of stage3] (output) {{Outputs}\\
{Daily Burned Masks \& Spatial Metrics}};;

\draw[arrow] (data) -- (stage1);
\draw[arrow] (data) -- (stage2);

\draw[arrow] (stage1.south) -- ++(0,0) |- (stage3.west);
\draw[arrow] (stage2.south) -- ++(0,0) |- (stage3.east);

\draw[arrow] (stage3) -- (output);

\end{tikzpicture}
\caption{\textcolor{black}{Overview of the three-stage wildfire modelling framework. 
Stage 1 estimates daily fire-occurrence probabilities, Stage 2 provides optional spread-rate quantile scenarios, and Stage 3 uses an RF-informed cellular automaton to simulate local spatial propagation.}}
\label{fig:pipeline_overview}
\end{figure}
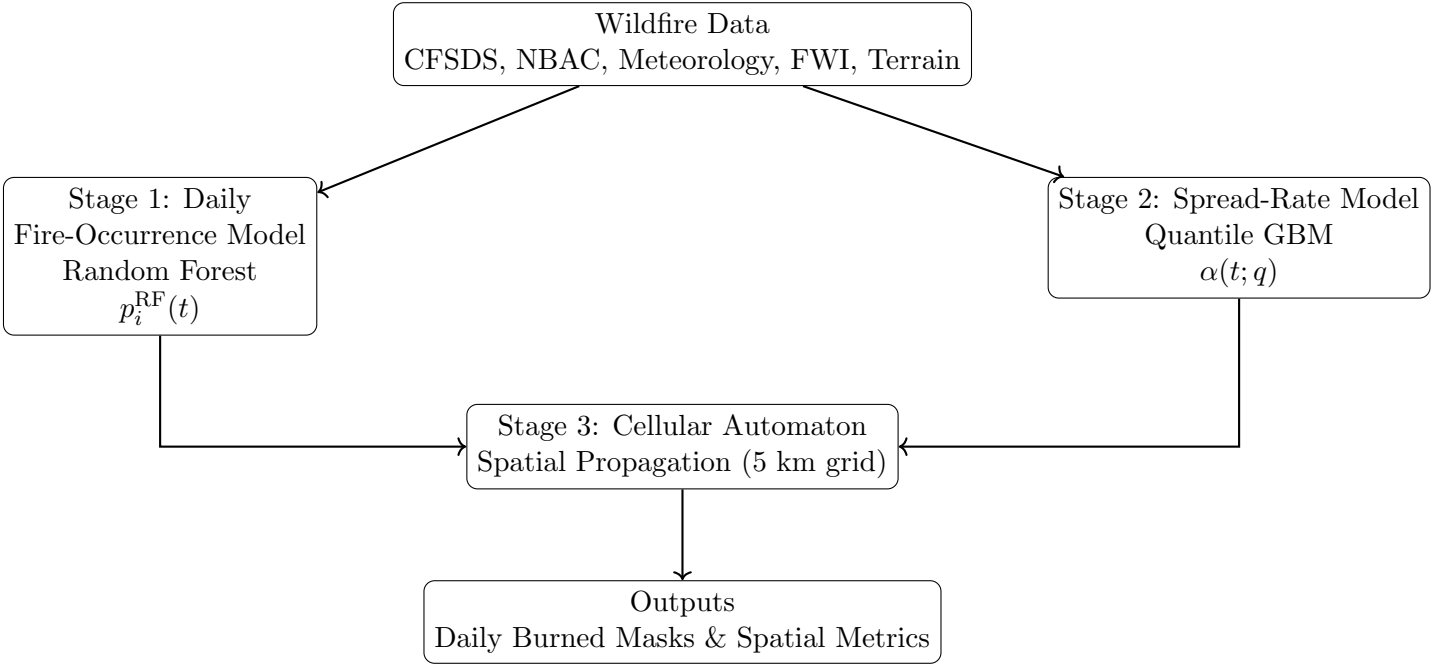

This study uses a three-stage modelling pipeline (Fig.~\ref{fig:pipeline_overview}) to estimate daily fire occurrence and simulate wildfire spread. 
\textcolor{black}{Stage~1 is a Random Forest classifier that estimates daily pixel-level fire-occurrence probabilities. 
Stage~2 uses a quantile gradient boosting model to estimate conditional spread-rate quantiles. 
Stage~3 is the main spatial simulation stage: it combines the RF probability layer with local neighbour-driven spread in an RF-informed CA.}

\textcolor{black}{This design separates three related but different tasks. 
Stage~1 estimates where burning is likely to be supported by environmental and landscape conditions. 
Stage~2 provides possible spread-strength inputs. 
Stage~3 combines these outputs in an RF-informed CA framework, where the RF layer provides daily spatial probabilities and the CA provides local neighbour-driven spread.}
\textcolor{black}{The Stage~1 and Stage~2 models are trained on data from the 2021 fire season and applied to later years without retraining. 
The Stage~3 simulations are mainly evaluated on the 2023 fire season, with selected 2024 diagnostics used to examine the RF probability layer.}
Because Stage~3 uses same-day meteorological inputs and an observed first-day fireline, the simulations should be interpreted as retrospective simulations rather than operational forecasts.
A graphical representation of the framework is given in Figure \ref{fig:pipeline_overview}.

\subsubsection{Stage~1: Random Forest Daily Fire-Occurrence Model}
The goal of Stage~1 is to predict whether a pixel is recorded as burning on
a given day. The binary response variable is defined as
$\mathbb{I}(\texttt{firearea}>0)$, where a positive label indicates that
burned area is recorded for the corresponding pixel-day observation.
Therefore, the response represents daily fire occurrence rather than the first ignition of a fire event. In the coupled framework, the resulting
RF probabilities are interpreted as a daily fire-occurrence probability layer rather
than as a direct simulation of fire propagation.  
The set of features in Stage~1 is limited to “static plus same-day dynamic” variables to avoid any leakage of future fire information. 
In total, 16 predictors were used which included static landscape attributes (elevation, slope, aspect, topographic wetness index, road density at 2\,km and 5\,km) and same-day weather and FWI indices (daily maximum temperature, noon relative humidity, noon wind speed, 24-hr precipitation, FFMC, DMC, DC, ISI, BUI, FWI). The predictors and their descriptions are summarized in Table~\ref{tab:predictors_used}. 
Variables that could potentially include realized fire activity or growth were excluded.

A Random Forest classifier 
was trained on the 2021 fire season data and applied to the 2022–2024 datasets without retraining.
The model was configured with 300 trees and \texttt{class\_weight} set to `balanced' to account for the imbalanced distribution of pixel-day labels (many non-burning observations and relatively few burning observations). 
At each split, the model randomly selected a subset of features, and bootstrap sampling was enabled. 
A fixed random seed was used for reproducibility. 
Missing values in the inputs were already imputed as described earlier.


To summarize, for each $i^{th}$ pixel, Stage~1 produces a daily fire-occurrence probability $p^{\mathrm{RF}}_i(t)$. 
In order to integrate these with the spread simulation, we aggregate pixel probabilities to a 5\,km grid i.e., for each cell $c$, we calculate the probability that at least one constituent pixel is burning on day $t$ using,
\[P^{\mathrm{RF}}_c(t) = 1 - \prod_{i \in c}\big(1 - p^{\mathrm{RF}}_i(t)\big)\,. \]

\textcolor{black}{The aggregated probability $P_c^{\mathrm{RF}}(t)$ is passed to Stage~3 as the RF probability layer.
It is not used to simulate fire spread by itself.
Instead, it is combined with the CA neighbour-spread probability.}
This aggregation assumes independence among the pixel-level probabilities. Because each 5\,km cell contains many spatially correlated pixels, the resulting coarse-grid probability may become large even when the individual probabilities are small.
\subsubsection{Stage~2: Quantile Gradient Boosting based Conditional Spread Model}

In Stage 2 we model the conditional distribution of a fire’s daily spread distance, measured in metres per day, given that a fire is already burning in a particular cell. 
Instead of predicting a single spread distance, we predict several quantiles ( i.e., the value below which a certain proportion of observations fall; for example, the 0.50 quantile is the median) to capture the variability in fire spread. 
We train a quantile gradient boosting model on the 2021 training data to estimate four quantiles $q = \{0.50,\,0.75,\,0.90,\,0.95\}$. These correspond to the 50th (median), 75th, 90th, and 95th percentiles for the distribution of the next-day fire spread distance. 
The median (50th quantile) represents a typical spread, while the higher quantiles represent increasingly extreme spread scenarios. 
\textcolor{black}{In this framework, the Stage~2 model is not used as a standalone burned-area simulator. 
Instead, its quantile outputs provide optional spread-rate scenarios for the Stage~3 RF-informed CA. 
In particular, the $q_{50}$ and $q_{75}$ outputs are used as spread-strength scenarios in the CA experiments.}
The response variable of the model trained in Stage~2 is the daily spread distance (field \texttt{sprdistm}).
The transformation $y = \log\!\big(1 + \texttt{sprdistm}\big)$ is applied to stabilize variance and improve the prediction of extreme values.
We trained separate gradient boosting models for each quantile $q$ using the same set of features. Because the models were fitted independently, quantile crossing could occur. We therefore report monotonicity violation rates in Section~\ref{sec:stage2_results}.

Stage~2 evaluates three spread-rate models trained on the 2021 dataset:
\begin{enumerate}
    \item A \textbf{Global} model without ecozone predictors, using the same set of meteorological, FWI, terrain, and human-accessibility variables listed in Table~\ref{tab:predictors_used}. The \textbf{Global} model serves as a baseline by evaluating predictive performance without explicit spatial or temporal spread information.
    
    \item An \textbf{ECO} model with one-hot ecozone encoding. This model is included to test whether explicit ecozone information improves spread-rate prediction, since different ecozones have different vegetation, climate, and terrain conditions.
    
    \item A \textbf{Global-PrevGrow} model, which extends the \textbf{Global} configuration with previous-day spread (\texttt{prevgrow}) as an additional dynamic predictor. This variable is used to test whether recent fire growth improves spread-rate prediction. It is also examined carefully because lagged spread information may introduce leakage if not handled consistently.
\end{enumerate}

\textcolor{black}{All models estimate the q50, q75, q90, and q95 quantiles of daily spread distance. The lower quantiles provide Stage 3 scenarios, while the upper quantiles assess extreme-spread behaviour.}
\textcolor{black}{The three configurations allow us to examine
three modelling assumptions: spatially homogeneous predictors (Global), explicit ecozone effects (ECO), and temporal spread dependence (Global-PrevGrow).}
Table~\ref{tab:stage2_combined} reports pinball loss,
empirical coverage for the [q75, q95] and [q90, q95] intervals,
and monotonicity violation rates for the [q90, q95] tail
across 2022--2024.

Similar to Stage 1, the predictor set for Stage~2 was chosen to avoid leakage of future information.
\textcolor{black}{To check the role of recent fire growth without introducing future information, we included one lagged variable, \textit{prevgrow}, which records the area burned on the previous day.
This variable is allowed because, at the start of day $t$, the burned area from day $t-1$ is already known.
In contrast, we excluded cumulative burned area and percent-growth variables, because they may encode the final fire size or later fire development.
This separation helps test whether lagged growth improves spread prediction while still avoiding information leakage.}

\textcolor{black}{Therefore, predictions from Stage~2 at time $t$ depend only on information available up to the beginning of day $t$ (including same day weather conditions but excluding any information from future fire growth).}
The outcome of Stage 2 is a set of daily spread-rate estimates, $\alpha(t;q)$, for each quantile $q$ and day $t$. These estimates are aggregated to the 5 km grid and used as optional dynamic spread-rate priors in Stage 3.
\textcolor{black}{Thus, Stage~2 supplies possible propagation-strength inputs for the CA. 
The final spatial spread simulation is carried out by the RF-informed CA in Stage~3.}

\subsubsection{Stage~3: RF-Informed Cellular Automaton for Spatial Propagation}
\label{sec:ca_method}

\textcolor{black}{In this coupled framework, the RF layer and the CA play different roles. 
The RF layer provides daily fire-occurrence probabilities based on environmental and landscape conditions. 
The CA then determines how fire spreads locally from burning cells to neighbouring cells. 
Therefore, the model 
is an RF-informed CA model: the RF layer provides a daily probability field, and the CA simulates local spread. The two components are combined probabilistically, so the RF layer does not act as a hard constraint on neighbour-driven spread.}
\textcolor{black}{The CA represents the landscape as a grid of cells and updates each cell's state based on local rules. 
In this study, the main CA sensitivity experiments use a 5\,km grid and daily time steps.}
\textcolor{black}{The parameter $\alpha$ is used to control the strength of neighbour-driven CA spread,
rather than as the main modelling target.}
Rather than aiming to exactly reproduce observed fire perimeters at high resolution, we focus on whether the model can generate a plausible envelope of fire spread for large fires under various scenarios.


As mentioned before, we choose a 5\,km grid because the daily spread distances are small relative to 5 km cells (order of 50–1000 m).
Thus, in one day the fire will typically spread at most to the adjacent cells. 
This reduces the mismatch between cell size and spread distance, thereby yielding more stable simulation behavior.
Recall that for each year we only considered fires that were large and occurred between DOB 150 - 250.
We further expand our region of interest by 0.1 degree in all directions to ensure a margin. 
All pixels inside this region are aggregated to 5 km cells using the majority burn status or any-burn to identify if a cell has burned on a given day.
We also create daily coarse-grid burned area maps from the CFSDS data for validation.


\subsubsection*{State system and update rules}

To define the update rules, we first define the state of each cell $c$ at day $t$ as $S_c(t) \in \{0,1,2\}$,
where 0 denotes \textit{unburned}, 1 denotes \textit{burning}, and 2 denotes \textit{burned}. The update rules for the CA model are given below.
\begin{itemize}
    \item \textbf{Burning to burned:} If a cell is burning at time $t-1$, then it becomes burned at time $t$.
    \item \textbf{RF probability component:} If a cell is unburned at time $t-1$, the RF model contributes the probability
$P^{\mathrm{RF}}_c(t)$ obtained from Stage~1.
 \textcolor{black}{This probability is derived from the daily fire-occurrence model at the coarse-grid level and excludes spread from neighbouring burning cells.}
\item \textbf{Neighbour driven spread:}
If at least one neighbour $n\in N(c)$ is burning at time $t{-}1$, the
probability of spread from $n$ to $c$ is
\[
p_{n\to c}(t)
= 1 - \exp\!\left(-\lambda \frac{\alpha_n(t)}{d}\right),
\]
\textcolor{black}{where $\alpha_n(t)$ is expressed in metres per day, $d=5000$\,m is the 5\,km coarse-cell size, and $\lambda$ is a scaling factor linking physical spread rate to probabilistic transition intensity.}
Assuming independence across neighbours, the overall neighbour-spread
probability is
\[
P^{\mathrm{spread}}_c(t)
= 1 - \prod_{n\in N(c)} \bigl[1 - p_{n\to c}(t)\bigr].
\]
\textcolor{black}{The exponential form converts the spread-strength input $\alpha_n(t)$ into a probability of spread from a burning cell to its neighbour. 
Larger values of $\alpha$ increase the local spread probability, while smaller values produce more conservative spread. 
Thus, $\alpha$ controls the strength of neighbour-driven spread in the CA.}
 
\textcolor{black}{These two mechanisms are combined to define the combined burning-transition probability used for state updates.}
\end{itemize}

\paragraph{Combined burning-transition probability.}
For an unburned cell, the combined burning-transition probability is
\[
P^{\mathrm{total}}_c(t)
= 1 - \Big(1 - P^{\mathrm{RF}}_c(t)\Big)\Big(1 - P^{\mathrm{spread}}_c(t)\Big).
\]
\textcolor{black}{This noisy-OR formulation combines the RF probability and the CA neighbour-spread probability. An unburned cell may become burning through either component. Therefore, the RF layer informs the CA simulation but does not prevent neighbour-driven spread when the RF probability is low.}
We then draw a Bernoulli outcome with probability $P^{\mathrm{total}}_c(t)$.
If the random draw is successful, we set $S_c(t)=1$ on day $t$.

\subsubsection*{Spread rate scenarios and experiments}
\textcolor{black}{We run the CA under multiple spread-strength scenarios. 
These include fixed $\alpha$ settings and quantile-based spread priors from Stage~2. 
The fixed settings are used to represent low, medium, and high propagation strengths. 
The quantile-based scenarios use the Stage~2 QGBM outputs as data-driven alternatives to fixed spread-strength settings.}

\textcolor{black}{We also include an $\alpha=0$ diagnostic case. 
When $\alpha=0$, neighbour-driven CA spread is removed, so the simulation isolates the role of the RF probability layer. 
This case is not treated as a competing spread model, because it cannot represent local fire propagation.}

\textcolor{black}{When $\alpha > 0$, the CA combines neighbour-driven spread with the RF probability layer.
By testing different $\alpha$ values, we examine how strongly this local spread should act.
If $\alpha$ is too large, the CA spread may move too far beyond the RF-based pattern.
Therefore, the comparison is a sensitivity analysis of spread strength, not a search for one best universal $\alpha$.}

\subsection{Evaluation metrics}

We compare simulated and observed burned masks up to each day $t$ and compute \textcolor{black}{true positives (TP), false positives (FP), false negatives (FN), and true negatives (TN)}.
We report the Intersection over Union (IoU) and Dice coefficient Dice$(t)$ which are defined below as,
\[
\mathrm{IoU}(t) = \frac{\mathrm{TP}}{\mathrm{TP} + \mathrm{FP} + \mathrm{FN}},
\qquad
\mathrm{Dice}(t) = \frac{2\,\mathrm{TP}}{2\,\mathrm{TP} + \mathrm{FP} + \mathrm{FN}}.
\]
We also report Cohen's $\kappa(t)$~\cite{Cohen1960}:
\[
\kappa(t) = \frac{p_o(t)-p_e(t)}{1-p_e(t)},
\]
where $p_o(t)$ is the actual match rate between the simulated and observed maps, and $p_e(t)$ is the match rate expected by chance.

Additionally, we also compare growth dynamics using cumulative burned area curves and
front-based distances. These include the Hausdorff distance between simulated
and observed perimeters, along with the mean and maximum front offsets. 
\textcolor{black}{For the CA simulations, Final IoU and Final Dice are computed at the end of the simulation period. 
Mean IoU and Mean Hausdorff Distance are averaged over the daily simulation horizon. 
Area Error is computed as the difference between simulated and observed burned area at the end of the season. 
These metrics allow us to evaluate both final burned-area overlap and daily spatial agreement during the simulated fire-growth process.}


\section{Results}

\subsection{Stage~1: Random Forest Daily Fire-Occurrence Model}
\label{sec:rf_results}

The Random Forest (RF) daily fire-occurrence classifier was trained on 2021 data using the \emph{static and safe dynamic} features 
and evaluated on the out-of-year datasets for years 2022–2024, restricted to the active fire season (Days of Burning 150–250). 
\textcolor{black}{Static predictors are variables that do not change with time,
such as terrain or ecozone.
Dynamic predictors are variables that change over time, such as temperature, humidity, and wind. We consider a variable to be ``safe" if it does not leak
future information. 
}
Each record was labeled as fire or non-fire based on the burned-area indicator $\mathbb{I}(\texttt{firearea}>0)$.
Table~\ref{tab:rf_crossyear} summarizes the cross-year results based on the metrics described below.
\begin{enumerate}
    \item \textbf{ROC curve (AUC):} The area under the ROC curve measures how well the model separates fire pixels from non-fire pixels. The ROC curve compares the true positive rate and the false positive rate at different probability thresholds. A higher AUC means that the model is better at distinguishing between fire and non-fire observations.
\item \textbf{Fire-class F1-score:} It combines precision and recall for the fire class, where precision measures how many predicted fire pixels are actually burning and recall measures how many true burning pixels are correctly detected by the model. The F1-score summarizes these two quantities and provides a single measure of fire detection performance.
\end{enumerate}
The model achieved AUC values of 0.725--0.795 across the three test years, indicating moderate and relatively stable discrimination on the 2022--2024 datasets.
\textcolor{black}{Baseline approaches in wildfire ignition modeling commonly include logistic regression and generalized linear models (GLMs), which provide simple and interpretable linear references\cite{Vilar2011,Chang2013}. Previous logistic-regression and GLM-based wildfire occurrence models have reported variable predictive performance across regions and model designs \cite{Chang2013,Vilar2011,Martin2019,Keeping2024DailyWildfireProbability,Pham2020ForestFireML}. In addition, operational fire risk assessment often relies on empirical indices such as the Fire Weather Index (FWI), which serves as a standard baseline in fire science \cite{VanWagner1987FWI,TaylorAlexander2021CFFDRS}. Compared to these approaches, machine learning models are often able to capture more complex nonlinear relationships and achieve improved predictive performance \cite{Jain2020Review,Ghorbanzadeh2019,Pang2022}.}
\vspace{0.5em}
\begin{table}[h]
\centering
\caption{Cross-year performance of the Random Forest daily fire-occurrence model
(trained on 2021, evaluated on 2022–2024, DOB 150–250).}
\label{tab:rf_crossyear}
\begin{tabular}{lcccc}
\hline
\textbf{Year} & \textbf{AUC} & \textbf{AP} & \textbf{F1 (fire)} & \textbf{Accuracy} \\
\hline
2022 & 0.795 & 0.903 & 0.879 & 0.812 \\
2023 & 0.728 & 0.931 & 0.844 & 0.750 \\
2024 & 0.725 & 0.951 & 0.910 & 0.844 \\
\hline
\end{tabular}
\label{tab:rf-stage1}
\end{table}
\textcolor{black}{These results support using the RF output as a daily probability layer in Stage~3. 
The RF model is not used as a standalone spread simulator. 
Instead, its daily probabilities are combined with the neighbour-spread component in the RF-informed CA.}




\subsection{Stage~2: Quantile Spread Models}
\label{sec:stage2_results}

\subsubsection{Model diagnostics}

Table~\ref{tab:stage2_combined} reports pinball loss,
empirical coverage for the [q75, q95] and [q90, q95] intervals,
and monotonicity violation rates for the [q90, q95] tail
across 2022--2024.

\textcolor{black}{
Pinball loss measures the accuracy of quantile predictions, while coverage
is the proportion of observations falling within the predicted quantile interval.
A monotonicity violation occurs when a higher predicted quantile is smaller
than a lower predicted quantile.
The nominal coverage rates of the [q75, q95] and [q90, q95] intervals are 0.20 and 0.05, respectively. Therefore, higher empirical coverage does not necessarily indicate better performance.
}
\begin{table}[H]
\centering
\caption{Comparison of three quantile spread models (Global, ECO, Global-PrevGrow)
across key extreme-spread diagnostics (2022--2024).
Pinball loss (lower is better) is reported for q90 and q95.
Coverage is the empirical proportion of observations falling inside the
[q75, q95] and [q90, q95] intervals, respectively.
``Viol.\,(90--95)'' denotes the monotonicity violation rate for the [q90, q95] pair.}
\label{tab:stage2_combined}
\renewcommand{\arraystretch}{1.15}
\begin{tabular}{l l c c c c c}
\toprule
\textbf{Year} & \textbf{Model} & \textbf{Quantile} &
\textbf{Pinball} &
\textbf{Cov$_{75,95}$} &
\textbf{Cov$_{90,95}$} &
\textbf{Viol.\,(90--95)} \\
\midrule

\multirow{6}{*}{\textbf{2022}}
& Global          & q90 & 0.1507 & \textbf{0.588} & 0.0615 & 0.149 \\
& ECO             & q90 & 0.1510 & 0.559 & 0.0227 & 0.558 \\
& Global-PrevGrow & q90 & \textbf{0.1489} & 0.454 & \textbf{0.0968} & \textbf{0.028} \\
& Global          & q95 & \textbf{0.0835} & \textbf{0.588} & 0.0615 & 0.149 \\
& ECO             & q95 & 0.0852 & 0.559 & 0.0227 & 0.558 \\
& Global-PrevGrow & q95 & 0.0889 & 0.454 & \textbf{0.0968} & \textbf{0.028} \\
\midrule

\multirow{6}{*}{\textbf{2023}}
& Global          & q90 & 0.2215 & \textbf{0.541} & 0.0520 & 0.154 \\
& ECO             & q90 & 0.1951 & 0.489 & 0.0088 & 0.777 \\
& Global-PrevGrow & q90 & \textbf{0.2274} & 0.448 & \textbf{0.127} & \textbf{0.0496} \\
& Global          & q95 & 0.1395 & \textbf{0.541} & 0.0520 & 0.154 \\
& ECO             & q95 & 0.1421 & 0.489 & 0.0088 & 0.777 \\
& Global-PrevGrow & q95 & \textbf{0.1377} & 0.448 & \textbf{0.127} & \textbf{0.0496} \\
\midrule

\multirow{6}{*}{\textbf{2024}}
& Global          & q90 & 0.1985 & \textbf{0.510} & 0.0253 & 0.207 \\
& ECO             & q90 & 0.1930 & 0.457 & 0.0134 & 0.677 \\
& Global-PrevGrow & q90 & \textbf{0.2022} & 0.342 & \textbf{0.0664} & \textbf{0.0494} \\
& Global          & q95 & 0.1213 & \textbf{0.510} & 0.0253 & 0.207 \\
& ECO             & q95 & \textbf{0.1202} & 0.457 & 0.0134 & 0.677 \\
& Global-PrevGrow & q95 & 0.1244 & 0.342 & \textbf{0.0664} & \textbf{0.0494} \\
\bottomrule
\end{tabular}
\end{table}

\subsubsection{Cross-year patterns in extreme-quantile behaviour}

Three clear patterns appear in Table~\ref{tab:stage2_combined}.

\paragraph{Global model: overcoverage for [q75, q95].}
For the wider interval [q75,q95], the Global model has the highest coverage in every year: 0.588 in 2022, 0.541 in 2023, and 0.510 in 2024. These values exceed the nominal level, suggesting overly wide intervals, although its pinball loss remains competitive in several cases (e.g., 0.0835 for q95 in 2022).

\paragraph{ECO model: low [q90, q95] coverage and many ordering violations.}
The ECO model has the lowest [q90, q95] coverage in all three years
(0.0227 in 2022, 0.0088 in 2023, 0.0134 in 2024).
It also has the highest violation rates (0.558--0.777).
In other words, ECO often predicts q95 below q90, and its [q90, q95] interval
contains relatively few observations.
In this setup, adding ecozone one-hot predictors does not improve the
upper-quantile results.

\paragraph{Global-PrevGrow model: higher [q90, q95] coverage with fewer violations.}
Global-PrevGrow has the highest [q90,q95] coverage in every year: 0.0968 in 2022, 0.127 in 2023, and 0.0664 in 2024. It also has the lowest violation rates (0.028–0.050), although its coverage remains above the nominal level.

\paragraph{Implications for Stage~3.}
The three models play different roles in the rest of the paper.
We retain Global as a simple baseline. Its higher coverage is not interpreted as better calibration.
We keep ECO only as a comparison case because it has low [q90, q95] coverage
and many violations.
We use the q50 and q75 outputs from Global-PrevGrow as dynamic spread scenarios in the RF-informed CA simulations. They are included for sensitivity analysis rather than as optimal priors, since Global-PrevGrow does not perform best on every metric.

\paragraph{Summary.}
Stage~2 shows clear differences among the three spread models.
Global shows overcoverage for [q75,q95], while Global-PrevGrow has fewer upper-quantile violations.
\textcolor{black}{The q50 and q75 outputs from Global-PrevGrow are used as optional spread-strength scenarios in Stage~3.}

\subsection{Stage 3: Cellular Automaton Spread Simulations}
\label{sec:ca_results}

This section reports the Stage~3 cellular automaton simulations. The main question is
how adding the RF probability layer changes the CA simulation relative to CA-only propagation. 
We first compare CA-only baselines with the
RF-informed CA model. We then examine how different $\alpha$ settings affect
neighbour-driven spread when combined with the RF layer.

Most results focus on the 2023 fire season on the 5\,km grid. 
The $\alpha$ comparison is a sensitivity test. It shows how different levels of local spread interact with the RF probability layer.

\textcolor{black}{The $\alpha=0$ case is not included in the main Results comparison because it removes neighbour-driven CA spread. 
It is treated only as an RF-layer diagnostic and is discussed separately in Section~5.}
Therefore, the main comparison does not directly quantify the incremental contribution of neighbour-driven spread beyond the RF layer.

\subsubsection{CA-only baselines and RF-informed CA}

We first compare CA-only spread models with the RF-informed CA model under the
same 5\,km grid and the same initial fireline. The CA-only baselines rely only on
local neighbour interactions. They do not use the RF probability layer, so
they have no additional data-driven probability component.

As shown in Table~\ref{tab:ca_rfca_comparison}, the deterministic CA-only model and
the probabilistic CA baseline recover only a small part of the observed burned
footprint. Their final IoU values are below 0.05, and both models strongly
underestimate burned area. In contrast, the RF-informed CA model gives much higher
spatial overlap. This shows that adding the RF probability layer substantially changes the simulated footprint at this coarse resolution.

\begin{table}[H]
\centering
\caption{End-of-season comparison between CA-only baselines and the RF-informed
CA model for the 2023 fire season at fixed spread intensity $\alpha=50$.
All models use the same 5\,km grid and the same initial fireline.}
\label{tab:ca_rfca_comparison}
\begin{tabular}{lccccc}
\toprule
Model & Final IoU & Mean IoU & Final Dice & Area Error (km$^2$) & HD (km) \\
\midrule
CA-only & 0.0327 & 0.1192 & 0.0632 & $-5925$ & 55.23 \\
Probabilistic CA & 0.0245 & 0.1012 & 0.0478 & $-5975$ & 58.52 \\
RF-informed CA & 0.9771 & 0.9649 & 0.9884 & 1500 & 94.00 \\
\bottomrule
\end{tabular}
\end{table}

The RF probability layer substantially changes the simulated burned footprint. The CA component retains local neighbour-driven spread. However, because the RF component can activate cells independently, this comparison does not by itself isolate the contribution of neighbour-driven CA spread.

The different $\alpha$ settings are then used to examine how the
strength of CA neighbour spread changes the RF-informed simulation.

We test fixed spread-rate priors
\[
\alpha \in \{50,100,300,1000,2000\},
\]
and quantile-based priors $q50$ and $q75$ from the Stage~2 QGBM model.
We treat $\alpha$ as a scenario-defining parameter rather than a directly observable
quantity. Instead of calibrating a single value, we select these $\alpha$ values to
represent low, medium, and high spread scenarios. These scenarios are informed by
empirical rate-of-spread studies and the range of predictions from the Stage~2
model. The purpose is to analyse model behaviour under different propagation
strengths, rather than to reconstruct any specific fire event.

Although the CA uses one propagation rule, changing $\alpha$ changes the neighbour
spread probability. In practice, the RF-informed CA results fall into three simple
spread levels: small ($\alpha=50,100$), medium ($\alpha=300, q50, q75$), and
stress-test ($\alpha=1000,2000$).

\subsubsection{Propagation rule and sensitivity groups}

Neighbour spread in the CA uses
\[
p = 1 - \exp(-\lambda \alpha / d),
\]
where $d=5$\,km is the grid spacing and $\lambda=0.1$ is a scale constant.
Here, $p(\alpha)$ is the daily chance that a burning cell ignites a neighbour.
With an eight-cell Moore neighbourhood, the expected number of newly ignited
neighbours per burning cell can be written as
\[
R(\alpha) = 8\, p(\alpha).
\]
This value gives a simple way to compare how strongly different $\alpha$ values
push the fire to expand through neighbours. The purpose of this comparison is not
to estimate a universal physical spread rate. Instead, it is used to examine how
much CA neighbour spread can be added before it starts to dominate or weaken the
RF-derived probability pattern.

\begin{table}[H]
\centering
\caption{Transition probability $p(\alpha)$ and expected number of ignited
neighbours $R(\alpha)=8p(\alpha)$ implied by the propagation rule.
Larger $\alpha$ values increase the strength of CA neighbour spread within the
RF-informed simulation.}
\label{tab:p_alpha}
\begin{tabular}{lccc}
\toprule
Regime & $\alpha$ (m/day) & $p(\alpha)$ & $R(\alpha)$ \\
\midrule
Small & 50  & 0.0010 & 0.008 \\
      & 100 & 0.0020 & 0.016 \\
\midrule
Medium & 300 & 0.0060 & 0.048 \\
       & $q50$ & $O(10^{-3}$--$10^{-2})$ & $O(0.02$--$0.07)$ \\
       & $q75$ & $O(10^{-3}$--$10^{-2})$ & $O(0.03$--$0.08)$ \\
\midrule
Stress Test & 1000 & 0.020 & 0.16 \\
            & 2000 & 0.040 & 0.32 \\
\bottomrule
\end{tabular}
\end{table}

\paragraph{Interpretation.}
Table~\ref{tab:p_alpha} shows how neighbour spread strength changes with $\alpha$.
For $\alpha=50$ and $\alpha=100$, $R(\alpha)$ is very small (below 0.02),
so spread through neighbours is limited.
For $\alpha=300$ and the quantile priors ($q50$ and $q75$), $R(\alpha)$ is larger
(around 0.05 in the fixed case), so spread is faster.
For $\alpha=1000$ and $\alpha=2000$, $R(\alpha)$ is much larger (0.16 to 0.32),
so neighbour ignition becomes much more likely.
As $\alpha$ increases, the neighbour-spread component has a greater influence on the final pattern.

We use these three groups to organize the RF-informed CA sensitivity results:
\[
\textbf{Small: } \alpha=50,100; \qquad
\textbf{Medium: } \alpha=300,\ q50,\ q75; \qquad
\textbf{Stress tests: } \alpha=1000,2000.
\]

\subsubsection{Small Regime: $\alpha=50,100$}

\paragraph{IoU Trajectories.}

\begin{figure}[H]
\centering
\includegraphics[width=0.80\linewidth]{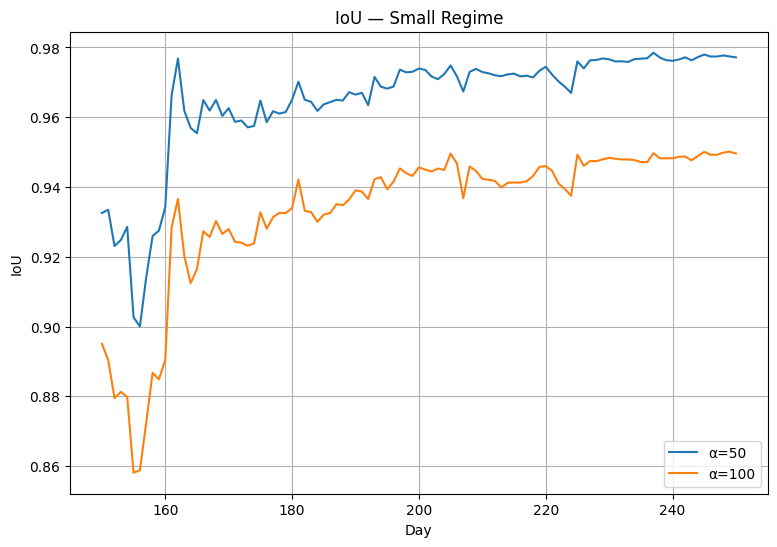}
\caption{Daily IoU for small-regime priors ($\alpha=50,100$).
\textcolor{black}{Both settings add limited neighbour spread when combined with the RF layer; $\alpha=50$
stays closest to the observed daily pattern among the nonzero settings.}}
\label{fig:iou_small}
\end{figure}

\textcolor{black}{In the small regime, the neighbour-spread contribution remains limited.
Only limited local growth is added around the RF-derived pattern.}
Both $\alpha=50$ and $\alpha=100$ follow the timing of daily spread.

Late in the season, $\alpha=100$ sometimes spreads more than observed,
so its IoU drops slightly compared with $\alpha=50$.
Overall, $\alpha=50$ stays closer to the observed daily pattern
throughout the season.
\textcolor{black}{This suggests that weak neighbour spread is more stable at the 5\,km scale,
rather than that $\alpha=50$ is a universal spread parameter.}

\paragraph{Cumulative Area Error.}

\begin{figure}[H]
\centering
\includegraphics[width=0.80\linewidth]{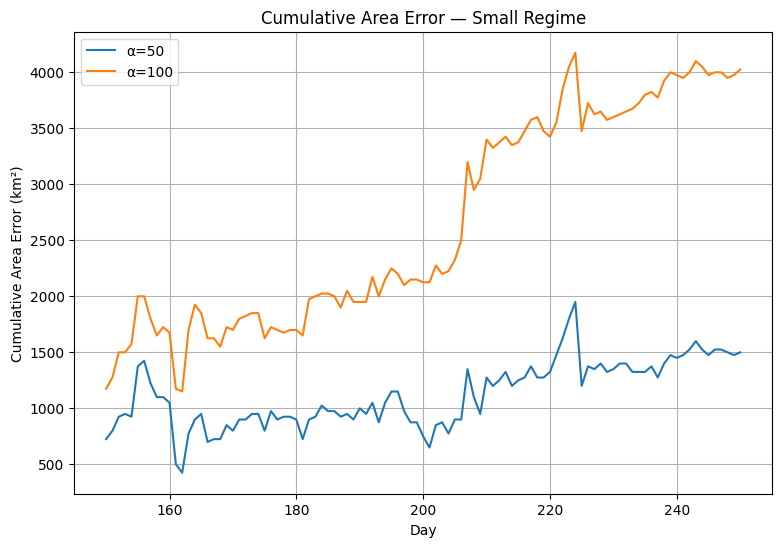}
\caption{Cumulative area error for small-regime priors.
$\alpha=50$ remains close to zero drift; $\alpha=100$ develops moderate
positive bias.}
\label{fig:area_small}
\end{figure}

Cumulative area error shows the same pattern.
\textcolor{black}{Weak CA spread adds local growth while staying close to the RF-derived pattern.}
By the end of the season, $\alpha=50$ has a final area error of about
1{,}500\,km$^2$, while $\alpha=100$ reaches about 4{,}025\,km$^2$
(Table~\ref{tab:alpha_sweep_2023_full}).
Compared with larger settings, these biases are much smaller.
This matches the higher spread probability for $\alpha=100$ in
Table~\ref{tab:p_alpha}.

\paragraph{Fireline Geometry.}

\begin{figure}[H]
\centering
\includegraphics[width=0.80\linewidth]{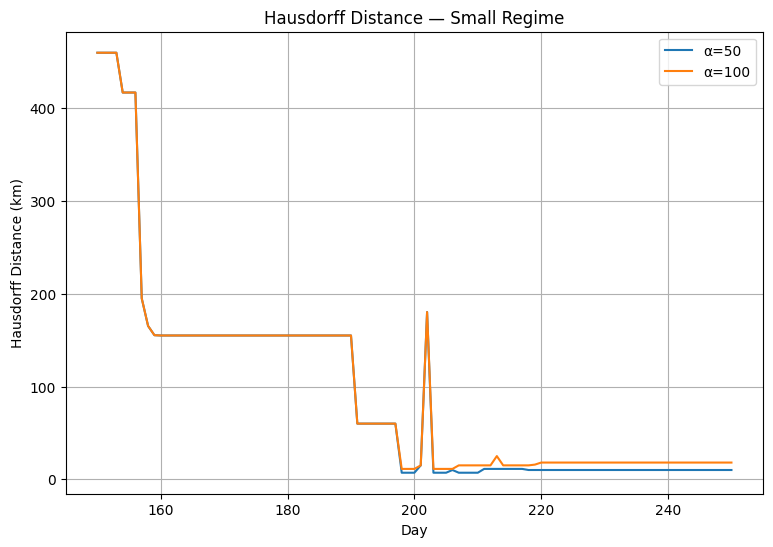}
\caption{Daily Hausdorff distance for small-regime priors.
\textcolor{black}{$\alpha=50$ gives the lowest geometric disagreement among the tested nonzero settings.}}
\label{fig:hd_small}
\end{figure}

Hausdorff distance measures how far the simulated and observed fire edges are.
\textcolor{black}{For $\alpha=50$, the fire edge remains close to the RF-derived spatial pattern.}
In this small group, $\alpha=50$ tends to give lower distances than $\alpha=100$.
A simple interpretation is that $\alpha=100$ spreads more evenly,
which can smooth the fire edge outward and move it away from the observed one.

\subsubsection{Medium Regime: $\alpha=300$, $q50$, $q75$}

\paragraph{IoU Behaviour.}

\begin{figure}[H]
\centering
\includegraphics[width=0.80\linewidth]{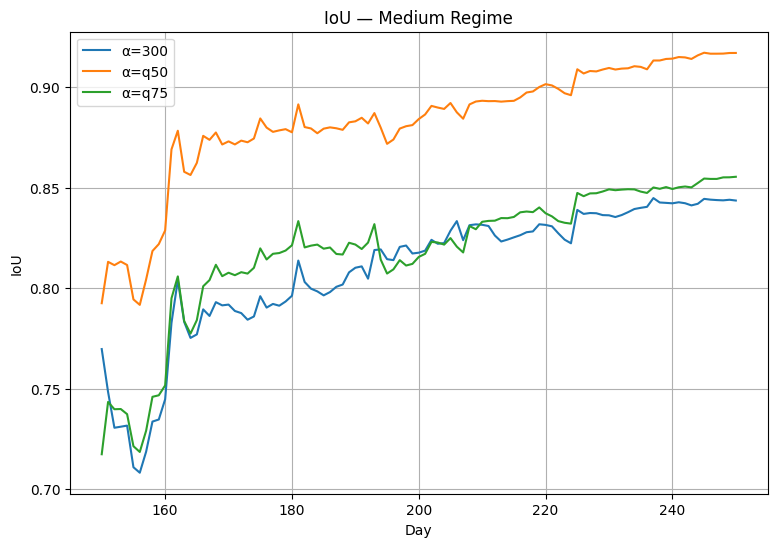}
\caption{Daily IoU for medium-regime settings. $q50$ and $q75$ closely
track $\alpha=300$, indicating similar CA spread strength when combined with the RF layer.}
\label{fig:iou_medium}
\end{figure}

\textcolor{black}{In the medium regime, CA spread starts to move beyond the RF-derived pattern.}
Compared with the small group, medium settings diverge earlier from the observed IoU.
The daily IoU curves for $q50$, $q75$, and $\alpha=300$ are close to each other,
which suggests that these choices lead to similar spread speed at this grid scale.

\paragraph{Cumulative Area Error.}

\begin{figure}[H]
\centering
\includegraphics[width=0.80\linewidth]{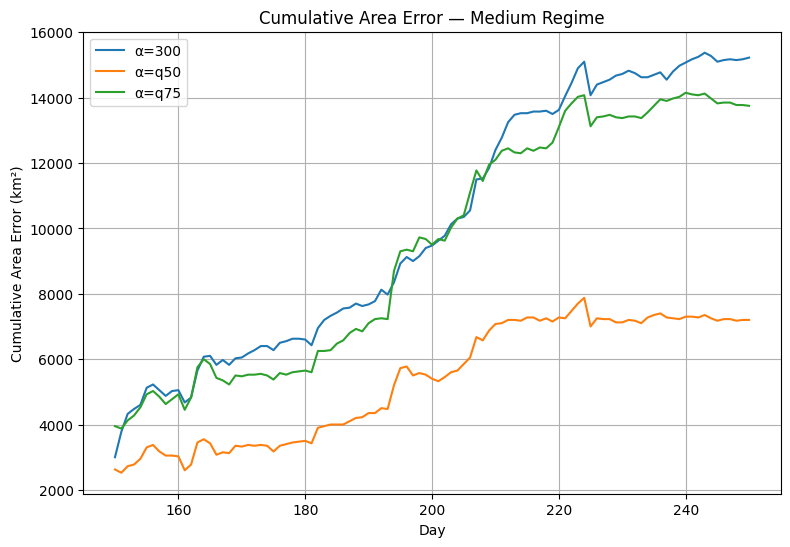}
\caption{Cumulative area error for medium-regime priors.
$q50$ and $q75$ finish between $\alpha=300$ and $\alpha=1000$ in area
bias.}
\label{fig:area_medium}
\end{figure}

All medium settings overestimate burned area by the end of the season.
\textcolor{black}{Stronger local spread moves the simulation farther from the RF-derived pattern
and leads to larger burned-area overestimation.}
From Table~\ref{tab:alpha_sweep_2023_full}, area error is 15{,}225\,km$^2$
for $\alpha=300$, 7{,}200\,km$^2$ for $q50$, and 13{,}750\,km$^2$ for $q75$.
These values are larger than the small group and much smaller than the stress tests.
\textcolor{black}{The q50 and q75 scenarios show that Stage~2 quantile outputs can be used as data-driven spread-rate priors in the CA. 
At the 5\,km grid, they behave similarly to medium fixed-spread settings. 
They are therefore better interpreted as alternative spread-strength scenarios rather
than as standalone spread simulations.}

\paragraph{Fireline Geometry.}

\begin{figure}[H]
\centering
\includegraphics[width=0.80\linewidth]{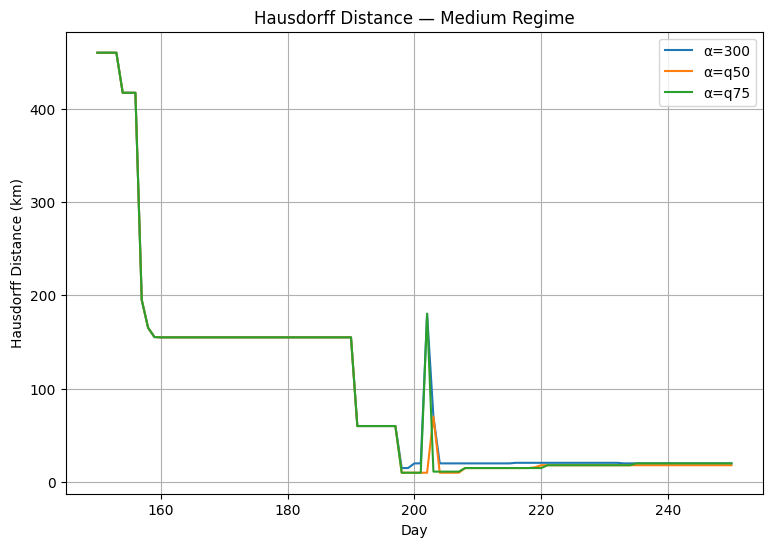}
\caption{Hausdorff distance for medium-regime priors.
All three settings exhibit larger and more persistent geometric
disagreement than the small regime.}
\label{fig:hd_medium}
\end{figure}

Medium settings also show larger and more persistent Hausdorff distances.

In simple terms, the simulated fire edges become smoother and more rounded,
and they drift outward compared with the observed edge.
Again, the quantile priors behave similarly to $\alpha=300$ in this case.

\subsubsection{Stress-Test Behaviour: $\alpha=1000,2000$}

\begin{figure}[H]
\centering
\includegraphics[width=0.80\linewidth]{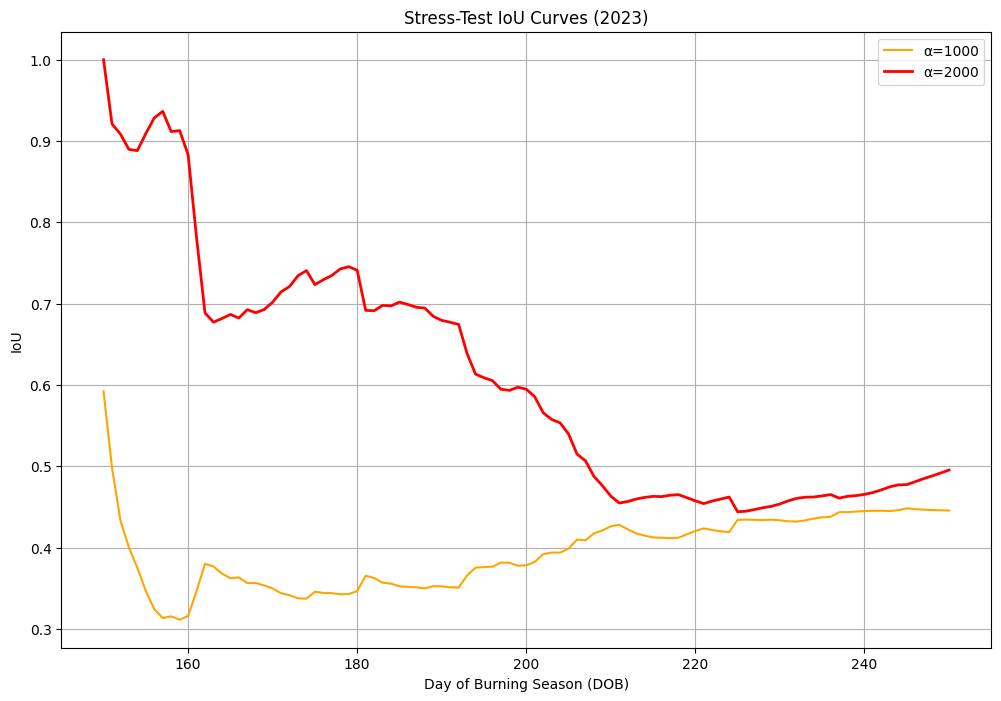}
\caption{Daily IoU for stress-test priors.
Both $\alpha=1000$ and $\alpha=2000$ diverge rapidly from observations.}
\label{fig:iou_stress}
\end{figure}

Stress-test settings spread far too fast at 5\,km scale.
\textcolor{black}{Here, neighbour-driven spread becomes too strong and dominates the combined update.}
IoU drops quickly, and area errors become extremely large.
From Table~\ref{tab:alpha_sweep_2023_full}, area error is about
104{,}000\,km$^2$ for $\alpha=1000$ and about 210{,}000\,km$^2$ for $\alpha=2000$.
These settings are therefore used only as upper-bound sensitivity tests.
\textcolor{black}{This confirms that very large $\alpha$ values can make neighbour-driven spread dominate the combined update, so
$\alpha$ should be interpreted as a propagation-strength control rather than as
the main target of the model.}

\subsubsection{End-of-season sensitivity comparison}

\begin{table}[H]
\centering
\caption{\textcolor{black}{End-of-season sensitivity metrics for RF-informed CA settings with neighbour-driven spread in 2023. 
Only $\alpha>0$ settings are included because the $\alpha=0$ case removes neighbour-driven spread and is treated separately as an RF-layer diagnostic. 
Among the nonzero settings, small propagation strengths produce smaller area bias and higher IoU than medium and stress-test settings. }}
\label{tab:alpha_sweep_2023_full}
\begin{tabular}{lccccc}
\toprule
\textcolor{black}{Propagation setting} & Final IoU & Mean IoU & Final Dice &
Area Error (km$^2$) & HD (km) \\
\midrule

50      & 0.9771 & 0.9649 & 0.9884 & 1500    & 94  \\
100     & 0.9496 & 0.9334 & 0.9741 & 4025    & 98  \\
300     & 0.8436 & 0.8092 & 0.9151 & 15225   & 100 \\
$q50$   & 0.9170 & 0.8827 & 0.9567 & 7200    & 96  \\
$q75$   & 0.8554 & 0.8180 & 0.9221 & 13750   & 98  \\
1000    & 0.4442 & 0.3936 & 0.6152 & 104000  & 121 \\
2000    & 0.1990 & 0.1720 & 0.3310 & 210000  & 144 \\
\bottomrule
\end{tabular}
\end{table}

\paragraph{Interpretation.}
\textcolor{black}{Table~\ref{tab:alpha_sweep_2023_full} summarizes the end-of-season sensitivity
results for 2023.}

\textcolor{black}{Among the $\alpha>0$ RF-informed CA settings, the stress-test cases
($\alpha=1000,2000$) have the lowest IoU and very large area errors. At this
level, the CA process largely dominates the combined update. Medium
settings ($\alpha=300$, $q50$, $q75$) sit in between. They have lower IoU and
larger positive area error than the small settings. Small settings
($\alpha=50,100$) keep the simulation closer to the RF-derived pattern while
adding limited neighbour growth.}

\textcolor{black}{Among the nonzero settings, $\alpha=50$ gives the most stable RF-informed CA
behaviour on the 5\,km grid. We therefore use it as the representative small-regime
setting in the qualitative example below.}

\subsubsection{High-resolution 500 m grid experiment}

\textcolor{black}{After the 5\,km sensitivity analysis, we further test the same RF-informed CA
structure on a finer 500\,m grid.}
 
\textcolor{black}{The CA rule still represented local neighbourhood spread, while the RF probability field contributed to the CA update.}
For this high-resolution test, a small fixed spread prior of
$\alpha=20$\,m/day was used.

\begin{figure}[H]
  \centering

  \begin{subfigure}[t]{0.31\linewidth}
    \centering
    \includegraphics[width=\linewidth]{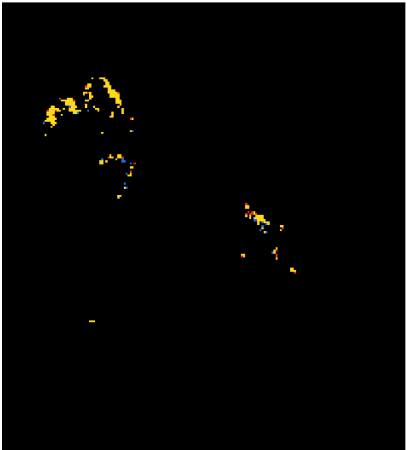}
    \caption{Case 1}
    \label{fig:rfca_500m_case1}
  \end{subfigure}
  \hfill
  \begin{subfigure}[t]{0.31\linewidth}
    \centering
    \includegraphics[width=\linewidth]{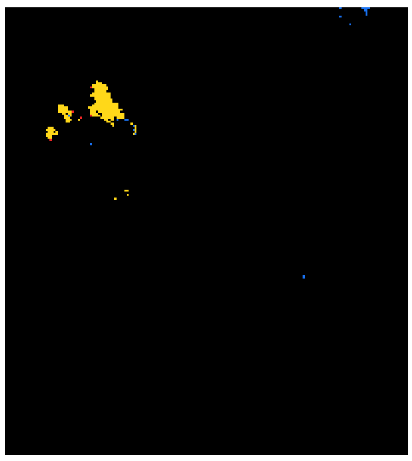}
    \caption{Case 2}
    \label{fig:rfca_500m_case2}
  \end{subfigure}
  \hfill
  \begin{subfigure}[t]{0.31\linewidth}
    \centering
    \includegraphics[width=\linewidth]{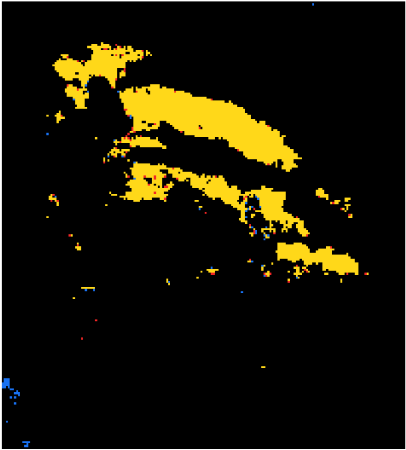}
    \caption{Case 3}
    \label{fig:rfca_500m_case3}
  \end{subfigure}

  \caption{Spatial error maps for the RF-informed CA simulation on the
  500\,m grid. Yellow indicates correct burned-area prediction, red indicates
  under-prediction, and blue indicates over-prediction. The examples show
  that the main burned-area footprint is largely captured, while most
  disagreement occurs near local fire-front boundaries and fragmented
  burned patches.}
  \label{fig:rfca_500m_error_maps}
\end{figure}

Figure~\ref{fig:rfca_500m_error_maps} shows three representative spatial
error maps on the 500\,m grid. Yellow cells indicate correct burned-area
prediction, red cells indicate under-prediction, and blue cells indicate
over-prediction. Across the three examples, correct predictions dominate
the main burned patches, suggesting that the RF-informed CA reproduces substantial portions of the broad burned-area footprint in these selected examples. The remaining red and blue cells
are mainly located near local boundaries or fragmented patches, indicating
that local fire-front placement remains uncertain.

This high-resolution experiment should be interpreted as a spatial
diagnostic rather than as a new parameter-optimisation result. The finer
grid provides more detailed local structure than the 5\,km grid, but it
also makes boundary-level disagreement more visible.

\subsubsection{Summary of CA Findings}

\textcolor{black}{The Stage~3 results show that CA-only propagation is not sufficient to recover
the large-scale burned footprint at the 5\,km scale. Without the RF-derived fire-occurrence probability layer, CA spread remains limited by the initial fireline and local neighbour rules.}

\textcolor{black}{Adding the RF probability layer changes the CA behaviour. The model
remains CA-based, but the CA update also includes a large-scale probability
pattern learned by the RF model. This keeps the local neighbour-spread mechanism
while reducing the dependence on the initial fireline alone.}

\textcolor{black}{Within the RF-informed CA settings, three spread regimes are observed. Small
settings ($\alpha=50,100$) keep CA spread close to the RF-derived pattern and
produce smaller area errors. Medium settings ($\alpha=300,q_{50},q_{75}$)
move farther from this pattern and lead to stronger area overestimation.
Stress-test settings ($\alpha=1000,2000$) make neighbour-driven spread dominate the combined update and
produce unrealistically fast spread. Among the $\alpha>0$ settings, $\alpha=50$
gives the most stable RF-informed CA behaviour on the 5\,km grid.}
\section{Discussion}
\label{sec:discussion}

\textcolor{black}{This section discusses the Stage~3 results in terms of the interaction between the RF probability layer and CA-based local propagation. 
The main question is not whether a single spread parameter can be optimized, but how the RF-informed CA behaves when neighbour-driven spread is combined with the RF probability layer. 
The discussion first compares CA-only baselines with the RF-informed CA, then interprets the role of $\alpha$, the use of Stage~2 quantile priors, and the effect of spatial resolution.}
\subsection{\textcolor{black}{Role of the RF Probability Layer}}
The RF layer helps the CA because it gives a data-driven map of where burning is more likely. 
The trained RF model also supports this interpretation. 
Fire weather and fuel-moisture variables, especially \texttt{ffmc}, \texttt{isi}, \texttt{fwi}, and \texttt{rh}, were among the most important predictors. 
This suggests that the RF output mainly reflects short-term fire weather conditions, while terrain and accessibility variables act as spatial modifiers. 
For this reason, the RF output is better understood as a daily fire-occurrence probability layer, not as a first-ignition model or a direct spread simulator.

\subsection{RF-layer diagnostic and the interpretation of $\alpha=0$}

The $\alpha=0$ case is used only as a diagnostic test.
When $\alpha=0$, the CA has no neighbour-driven spread.
Burning transitions therefore arise only from the RF probability layer.
This helps show how much of the 5 km burned footprint can be explained by the RF layer alone.

However, this does not make $\alpha=0$ a real spread model.
Without neighbour spread, the model cannot describe how fire grows from one burning cell to nearby cells.
Therefore, $\alpha=0$ is kept only to explain the role of the RF layer.
Because the $\alpha=0$ result is not included in the main quantitative comparison, the current analysis does not fully isolate the contribution of neighbour-driven CA spread.

The main analysis focuses on $\alpha>0$, where the CA combines local spread with the RF probability layer.

At 5 km resolution, the RF layer may already explain much of the broad burned footprint. Therefore, final IoU alone is not enough to evaluate spread skill. Daily growth, area error, and fire-front geometry should also be considered.

\subsection{Neighbour-driven CA baselines at coarse resolution}

Classical CA-based fire spread models rely on local neighbour interactions
to propagate fire across a spatial grid. When applied at coarse
resolution, such models are expected to capture broad fire growth trends
but often struggle to reproduce the full spatial extent of large wildfire
events. This limitation is evident in the CA-only baseline evaluated here,
which produces very low spatial overlap with the observed burned area.
Despite sustained neighbour-driven expansion, the simulated fire remains
confined to a small region around the ignition zone, resulting in a
severe underestimation of total burned area.

Introducing stochasticity through an ensemble-based probabilistic CA does
not fundamentally alter this behaviour. Although the probabilistic CA
accounts for randomness in local spread decisions and yields a burn
probability map, its deterministic envelope remains spatially constrained
by the same neighbour interaction rules. As a result, the probabilistic CA
baseline exhibits overlap and area errors comparable to those of the
deterministic CA-only model.

\subsection{Quantitative comparison under matched conditions}
To illustrate these differences quantitatively,
Table~\ref{tab:ca_rfca_comparison}
summarises end-of-season metrics for the 2023 fire under a fixed spread
intensity $\alpha=50$. All models are initialised using the observed
first-day fireline and evaluated on the same 5\,km grid.

The contrast between the CA baselines and the RF-informed CA framework is
substantial. Both CA-only and probabilistic CA achieve final IoU values
below 0.05 and underestimate the burned area by nearly 6000~km$^2$,
indicating that neighbour-driven propagation alone is insufficient to
recover the spatial footprint of the 2023 fire. In contrast, RF-informed CA
achieves near-complete spatial overlap with the observed burn scar, with a
final IoU of 0.9771 and a Dice coefficient of 0.9884.

\subsection{\textcolor{black}{Role of propagation strength and quantile priors}}

The sensitivity results show that $\alpha$ should be interpreted as a spread-strength control, not as a universal fire-spread parameter.
Small values such as $\alpha=50$ and $\alpha=100$ keep CA spread close to the RF-derived probability pattern.
Medium values, including $\alpha=300$, q50, and q75, allow more local spread and lead to stronger area overestimation.
Very large values such as $\alpha=1000$ and $\alpha=2000$ increase the influence of neighbour-driven spread and produce unrealistically fast spread.

The q50 and q75 scenarios show how Stage 2 outputs can be used inside the RF-informed CA.
They are not standalone spread simulations.
At the 5 km grid, they mainly act as alternative spread-strength inputs for the CA update.

\subsection{Why RF-informed CA behaves differently}

The different behaviour of RF-informed CA reflects the addition of the RF probability component to local CA spread. The Random Forest learns a statistical representation of daily fire occurrence from historical data, while the CA provides spatial continuity through neighbour transitions. Because either component can activate an unburned cell, the current comparison does not fully isolate the contribution of local CA propagation.

This separation of roles provides a simple and interpretable way to combine a data-driven probability field with spatially explicit propagation.

\subsection{Interpreting geometric error metrics}

Although RF-informed CA exhibits a larger Hausdorff distance than the CA baselines
(Table~\ref{tab:ca_rfca_comparison}), it also achieves substantially higher overall spatial overlap. A larger Hausdorff distance indicates a greater maximum boundary error. Therefore, the RF-informed CA boundary agreement is worse than the IoU result alone would suggest. IoU and Hausdorff distance measure different aspects of spatial performance and should be interpreted together.

\subsection{Implications for baseline comparison}

Under matched spread intensity and identical initial conditions, adding the RF probability layer substantially changes the simulated burned footprint relative to the evaluated CA-only baselines. This produces substantially higher spatial overlap in the reported 2023 experiment. However, the present comparison does not establish how much of this improvement is contributed by the CA relative to the RF layer alone.

\section{Conclusion}
This study developed an RF-informed Cellular Automaton framework for large-scale wildfire spread modelling.
The RF model provides a daily fire-occurrence probability layer, while the CA adds local neighbour-driven spread. The two components are combined probabilistically rather than through a hard spatial constraint.
This design separates the question of where burning is likely to occur from the question of how fire spreads locally.

The results show that the framework should not be interpreted as a search for a
universal optimal $\alpha$. Instead, $\alpha$ controls the strength of neighbour-driven CA spread when combined with the RF probability layer. At the 5~km grid, small nonzero $\alpha$
values gave the most stable results. In particular, $\alpha = 50$ performed best
among the tested settings. This value is specific to this grid and model setup.
Larger values increased the influence of neighbour-driven spread and led to stronger area overestimation.

The Stage 2 scenarios did not consistently outperform the fixed settings. They are therefore treated as sensitivity cases rather than evidence of improvement.

The main Stage~3 results are limited to the 2023 fire season, so the cross-year performance of the full RF-informed CA has not yet been established. The current comparison does not fully separate the contribution of neighbour-driven CA spread from that of the RF layer, and the larger Hausdorff distance indicates remaining boundary errors. Because same-day meteorological inputs were used, the simulations should be interpreted as retrospective rather than operational forecasts.

These findings also point to the importance of grid resolution. 
The stable behaviour observed at the 5 km scale should be interpreted within the spatial scale used in this study. 
At finer resolutions, the meaning of one neighbour transition changes, and the same value of \(\alpha\) may lead to different propagation behaviour. 
The 500\,m diagnostic experiment shows that a finer grid can reveal more detailed local boundary structure, while also making fire-front disagreement more visible. 

Future work should evaluate the RF-only baseline directly, examine the coarse-grid probability aggregation, and validate Stage~3 across additional years and fire events.

\appendix

\section{Supplementary Material}

\subsection{Predictors used in the models}

\begin{table}[H]
\small
\centering
\caption{Predictors used in Stage 1 and Stage 2 models.}
\label{tab:predictors_used}
\begin{tabular}{lllll}
\hline
Category & Variable & Description & Temporal type & Stages\\
\hline
Meteorology
& tmax & Maximum daily temperature (C) & Dynamic (same day) & 1,2\\
& rh & Noon relative humidity (\%) & Dynamic (same day) & 1,2 \\
& ws & Noon wind speed at 10 m (km/h) & Dynamic (same day) & 1,2 \\
& prec & 24 hr precipitation (mm) & Dynamic (same day) & 1,2 \\
FWI system
& ffmc & Fine Fuel Moisture Code & Dynamic (same day) & 1,2 \\
& dmc & Duff Moisture Code & Dynamic (same day) & 1,2 \\
& dc & Drought Code & Dynamic (same day) & 1,2 \\
& isi & Initial Spread Index & Dynamic (same day) & 1,2 \\
& bui & Buildup Index & Dynamic (same day) & 1,2 \\
& fwi & Fire Weather Index & Dynamic (same day) & 1,2 \\
Terrain
& dem & Elevation (m) & Static & 1,2 \\
& slope & Slope (degrees) & Static & 1,2 \\
& aspect & Aspect (degrees) & Static & 1,2 \\
& twi & Topographic wetness index & Static & 1,2 \\
Human access
& roaddens2k & Road density within 2 km (km/km$^2$) & Static & 1,2 \\
& roaddens5k & Road density within 5 km (km/km$^2$) & Static & 1,2 \\
Lagged growth
& prevgrow & Fire growth on prior fire day (ha) & Dynamic (lagged) & 2 \\
\hline
\end{tabular}
\end{table}

\subsection{Variables excluded to avoid information leakage}
\begin{table}[H]
\centering
\caption{Key fire growth fields not used as predictors.}
\label{tab:excluded_leakage}
\begin{tabular}{llll}
\hline
Field & Meaning & Role in this study & Reason \\
\hline
firearea & Daily burned area (ha) & Label only in Stage 1 & Realised growth \\
cumuarea & Cumulative burned area (ha) & Not used & Encodes realised spread \\
pctgrowth & Percent growth this day (\%) & Not used & Encodes realised spread \\
sprdistm & Daily spread distance (m) & Target in Stage 2 & Response variable \\
\hline
\end{tabular}
\end{table}

\subsection{Construction of the CFSDS burn progression data}
The CFSDS pixel-level daily fire progression data were constructed from active-fire detections from the Moderate Resolution Imaging Spectroradiometer (MODIS) and the Visible Infrared Imaging Radiometer Suite (VIIRS), accessed through NASA FIRMS \cite{Barber2024CFSDS,NASAFIRMS2026}. 
FIRMS provides near-real-time active-fire data from both MODIS and VIIRS, while the MODIS and VIIRS products provide active-fire or thermal-anomaly detections at different spatial resolutions \cite{NASAFIRMS2026,NASAMODISFire2026,NASAVIIRSFire2026}.
In CFSDS, the day of burning (DOB) for each pixel was estimated from MODIS and VIIRS active-fire detections using ordinary kriging on a 180\,m grid \cite{Barber2024CFSDS}. 
The resulting DOB raster was then masked to burned areas identified by the National Burned Area Composite (NBAC), producing a day-by-day burn progression map for each fire \cite{Barber2024CFSDS}.

\subsection{Definitions of Evaluation Metrics}
\label{sec:metrics}

This section provides mathematical definitions of all evaluation metrics used in the main text.

\paragraph{Intersection over Union (IoU).}
For predicted burned area $A$ and observed burned area $B$,
\[
\mathrm{IoU} = \frac{|A \cap B|}{|A \cup B|}.
\]
IoU measures the spatial overlap between simulated and observed burned areas.

\paragraph{Dice score.}
\[
\mathrm{Dice} = \frac{2|A \cap B|}{|A| + |B|}.
\]
Dice score is another overlap metric, placing more weight on the shared burned area.

\paragraph{Hausdorff distance.}
Let $\partial A$ and $\partial B$ denote the boundaries of the predicted and observed burned areas. The Hausdorff distance is defined as
\[
d_H(\partial A,\partial B) =
\max\left\{
\sup_{x \in \partial A} \inf_{y \in \partial B} d(x,y),
\sup_{y \in \partial B} \inf_{x \in \partial A} d(x,y)
\right\}.
\]
It measures the largest geometric discrepancy between simulated and observed fire perimeters.

\paragraph{Maximum distance error.}
The maximum distance error is defined as the maximum pointwise distance between predicted and observed fire perimeters.
It provides a simple measure of the worst-case spatial deviation.

\paragraph{Pinball loss.}
For quantile level $\tau \in (0,1)$, the pinball loss is defined as
\[
L_{\tau}(y,\hat{y}) =
\max\big(\tau(y-\hat{y}), (\tau-1)(y-\hat{y})\big).
\]
It is used to evaluate quantile predictions, penalizing under- and over-estimation asymmetrically.

\subsection{Additional Metrics and Supporting Results}
\label{sec:additional_results}

This section reports additional metrics and figures that support the main results but are not included in the main text for clarity.

\subsubsection{Cumulative Area Error}

Cumulative area error measures the difference between simulated and observed burned area over time.
While informative, it is not used as a primary metric in the main text, where we focus on IoU and Hausdorff distance.


\subsubsection{Additional Distance-Based Metrics}

We also evaluated additional distance-based metrics (e.g., maximum distance error) to assess extreme deviations.
These results are consistent with the Hausdorff distance trends reported in the main text.

\clearpage
\addcontentsline{toc}{section}{References} 

\bibliographystyle{apalike}
\bibliography{references}

\end{document}